\documentstyle[amssymb,prl,aps,psfig,epsf]{revtex} 
 
\begin{document} 
\draft 
\title{Giant thermoemf in multiterminal superconductor/normal metal mesoscopic 
structures.} 
\author{R. Seviour$^{\ast }$ and A.F. Volkov$^{\ast \dagger }$} 
\address{$^*$ School of Physics and Chemistry,\\ 
Lancaster University, Lancaster LA1 4YB, U.K.\\ 
$^{\dagger}$Institute of Radioengineering and Electronics of the Russian \\ 
Academy of Sciencies, Mokhovaya str.11, Moscow 103907, Russia.} 
\date{\today} 
\maketitle 
 
\begin{abstract} 
We considered a mesoscopic superconductor/normal metal (S/N) structure in 
which the N reservoirs are maintained at different temperatures. It is shown 
that in the absence of current between the N reservoirs a voltage difference  
$V_{T}$ arises between the superconducting and normal conductors. The 
voltage $V_{T}$ oscillates with increasing phase difference $\varphi $ 
between the superconductors, and its magnitude does not depend on the small 
parameter $(T/\epsilon _{F}).$ 
\end{abstract} 
 
\smallskip

It is well known that if the terminals of a normal conductor are maintained 
at different temperatures, then in the absence of a current a thermoelectric 
voltage ($V_{emf})$ appears between the terminals. The magnitude of $V_{emf}$ 
is equal to $c_{1}(T/\epsilon _{F})\delta T/e,$ where $c_{1}$ is a factor of 
the order 1, $\epsilon _{F}$ is the Fermi energy and $\delta T$ is the 
temperature difference (see for example \cite{Abr}). 
 
In this paper we analyse the thermoelectric effect in the mesoscopic 
structure shown in Fig.1. We show that when a temperature difference exists 
between the normal (N) reservoirs, a voltage between normal and 
superconducting circuits $V_{T}$ appears. Unlike in the normal case, the 
magnitude of this voltage does not depend on a small parameter $(T/\epsilon 
_{F})$, and also this voltage oscillates as the phase difference $\varphi $ 
between the superconductors varies. We assume that the superconductors are 
connected via a superconducting loop and the phase difference between them $%
\varphi $ is controlled by an applied magnetic field. There is no current 
between the N reservoirs and the temperatures of the reservoirs are 
different: $T(\pm L)=T_{o}\pm \delta T$. We will calculate the electric 
potential in the N film and, in particular, the potential $V_{T}$ in the N 
reservoirs. Since we set the potential in the superconductors equal to zero, 
the potential $V_{T}$ is the voltage difference between the N reservoirs and 
superconductors which arises in the presence of the temperature difference $%
\delta T$. 
 
In order to find the potential $V_{T}$, we need to determine the 
distribution functions $f_{\pm }$ and the condensate wave functions $%
\widehat{F}^{R(A)}$ induced in the N film. The distribution functions $%
f_{\pm }$ are related to the ordinary distribution functions for electrons $%
n_{\uparrow }$ and holes $p_{\downarrow }$: $f_{+\uparrow }=f_{+\downarrow 
}=1-(n_{\uparrow }+$ $p_{\downarrow })$ and $f_{-\uparrow }=f_{-\downarrow 
}=-(n_{\uparrow }-$ $p_{\downarrow })$ (we assume that there is no 
spin-dependent interaction in the system). The function $f_{+}$ determines 
the condensate current and the function $f_{-}$ determines the quasiparticle 
current and electrical potential \ (see for example Ref. \cite{LO},where $%
f_{+}$ and $f_{-}$ are denoted by $f$ and $f_{1\text{ }}$respectively, and 
Ref. \cite{LRaim} where the application of the Green's function technique to 
the study of transport in S/N mesoscopic structures is discussed). These 
functions satisfy the kinetic equation (see Ref. \cite{SV}) 
 
\begin{equation} 
L{\Large \partial }_{x}[{\Large M}_{\pm }{\Large \partial }_{x}f_{\pm }(x)%
{\Large +J}_{S}f_{\mp }(x)\pm {\Large J}_{an}{\Large \partial }_{x}f_{\mp 
}(x)]{\Large =r[}\stackrel{}{A_{\pm }}{\Large \delta (x-L}_{1}{\Large )+}%
\stackrel{\_}{A_{\pm }}{\Large \delta (x+L}_{1}{\Large )].} 
\end{equation} 
where all the coefficients are expressed \ in terms of the retarded 
(advanced) Green's functions: $\widehat{G}^{R(A)}=G^{R(A)}\hat{\sigma}_{z}+%
\widehat{F}^{R(A)};$ $\ M_{\pm }=(1-G^{R}G^{A}\mp (\widehat{F}^{R}\widehat{F}%
^{A})_{1})/2;$ $J_{an}=(\widehat{F}^{R}\widehat{F}^{A})_{z}/2,\ J_{s}=(1/2)(%
\widehat{F}^{R}\partial _{x}\widehat{F}^{R}-\widehat{F}^{A}\partial _{x}%
\widehat{F}^{A})_{z},$\ $A_{\pm }=(\nu \nu _{S}+g_{1\mp })(f_{\pm }-f_{S\pm 
})-(g_{z\pm }f_{S\mp }+g_{z\mp }f_{\mp });$ $g_{1\pm }=(1/4)[(\widehat{F}%
^{R}\pm \widehat{F}^{A})(\widehat{F}_{S}^{R}\pm \widehat{F}_{S}^{R})]_{1};$ $%
g_{z\pm }=(1/4)[(\widehat{F}^{R}\mp \widehat{F}^{A})(\widehat{F}_{S}^{R}\pm  
\widehat{F}_{S}^{R})]_{z}.$ The parameter $r=R/R_{b}$ is the ratio of the 
resistance of the N wire $R$ and S/N interface resistance $R_{b}$; the 
functions $\stackrel{\_}{A_{-}}$ and $\stackrel{\_}{A}_{+}$ coincide with $%
\stackrel{}{A_{-}},\stackrel{}{A}_{+}$ if we make a substitution $\varphi 
\rightarrow -\varphi $. We introduced above the following notations $(%
\widehat{F}^{R}\widehat{F}^{A})_{1}=Tr(\widehat{F}^{R}\widehat{F}^{A}),$ $(%
\widehat{F}^{R}\widehat{F}^{A})_{z}=Tr(\hat{\sigma}_{z}\widehat{F}^{R}%
\widehat{F}^{A})$ etc.; $\nu ,$ $\nu _{S}$ are the density-of states in the 
N film at $x=L_{1}$ and in the superconductors. The functions $f_{S\pm }$ 
are the distribution functions in the superconductors which are assumed to 
have equilibrium forms. This means that $f_{S+}\equiv f_{eq}=\tanh (\epsilon 
\beta _{o})$ and $f_{S-}=0,$ because we set the potential of the 
superconductors equal to zero (no branch imbalance in the superconductors).  
 At the reservoirs the distribution functions $f_{\pm }$ obey the 
boundary conditions: $f_{\pm }(\pm L)=F_{V\pm }(\pm L),$  where $%
F_{V\pm }(L)=(1/2)[\tanh (\beta _{o}(\epsilon +eV_{T}))\pm \tanh (\beta 
_{o}(\epsilon -eV_{T}))],\beta _{o}=(2T_{o})^{-1}.$ 
 
In order to clarify the physical meaning of different terms in Eq.(1), 
consider the equation for $f_{-}$. The first term in this equation is the 
partial quasiparticle current (the quasiparticle current at a given energy) 
in the N film . The second term is the condensate current and the third term 
is also the condensate current which appears under nonequilibrium 
conditions. The factors $A_{-}$ and $\stackrel{\_}{A_{-}}$ are the partial 
currents through the S/N interfaces. The first term in $A_{-}$ is the 
partial quasiparticle current above (the term $\nu \nu _{S}f_{-}$) and below 
(the term $g_{1\mp }f_{-}$) the gap $\Delta ,$ and the last term is the 
condensate current. The factors $A_{+}$ and $\stackrel{\_}{A_{+}}$ are equal 
to zero below the gap (complete Andreev reflection). 
 
In this paper we consider the case when the S/N interface resistance is 
greater than or approximately equal to the resistance of the N wire ($r<1$). 
However $r$ should not be too small because in Eq.(1) we neglected the 
inelastic collision integral. This implies that the condition 
 
\begin{equation} 
(\epsilon \tau _{\epsilon })^{-1}<<r\lesssim 1 
\end{equation} 
must be fulfilled; here $\tau _{\epsilon }$ is the energy relaxation time, $%
\epsilon \approx \min \{T,\epsilon _{L}\},$ and $\epsilon _{L}=D/L^{2}$ is 
the Thouless energy. We consider the most interesting case of low 
temperatures ($T<<\Delta $) when for characteristic energies $\epsilon 
<<\Delta $ the functions $A_{+}$ and{\Large \ }$\stackrel{\_}{A_{+}}$ equal 
zero as in this case $\nu _{S}\approx 0$ and $\widehat{F}^{R}\cong \widehat{F%
}^{A}.$ The first integration of Eq.(1) yields 
 
\begin{equation} 
{\Large M}_{+}{\Large \partial }_{x}f_{+}(x){\Large +J}_{S}f_{-}(x)+{\Large J%
}_{an}{\Large \partial }_{x}f_{-}(x){\Large =J}_{+} 
\end{equation} 
where $J_{+}$ is a constant of integration. In the limit of the small 
parameter $r$, all the corrections related to the proximity effect are small 
(the functions $\widehat{F}^{R(A)}\stackrel{}{\text{are proportional to }r%
\text{). Therefore}}$ in the main approximation in the parameter $r,$ we 
have for the distribution function $f_{+}(x)$ 
 
\begin{equation} 
f_{+}(x)\cong \delta \beta \epsilon (x/L)\cosh ^{-2}(\beta _{o}\epsilon 
)+f_{eq} 
\end{equation} 
 
In obtaining Eq.(4) we have taken into account that in the N reservoirs the 
function $f_{+}(\pm L)$\ has an equilibrium form with different temperatures:%
 $f_{+}(\pm L)=\allowbreak f_{eq}\pm \delta \beta \epsilon \cosh 
^{-2}(\beta _{o}\epsilon ),$ where $\delta \beta /\beta \equiv 
-\delta T/T,\beta _{o}=(2T_{o})^{-1}$.  Eq.(4) implies that the 
temperature gradient leads to a flow of nonequilibrium electrons and holes 
in the N wire. In order to find the distribution function $f_{-}(x)$, we 
integrate Eq.(1) 
 
\begin{equation} 
{\Large M}_{-}{\Large \partial }_{x}\stackrel{}{f_{-}(x)}{\Large +J}_{S}%
{\Large f}_{+}(x){\Large -J}_{an}{\Large \partial }_{x}f_{+}(x){\Large =J}%
_{1}\Theta (L_{1}-\mid x\mid )+{\Large J}_{2+,2-}\Theta (\mid x\mid -L_{1}) 
\end{equation} 
 
 The constants $J_{1}$ and $J_{2+,2-}$ are the partial 
currents in regions $(-L_{1},+L_{1})$ , $(L_{1},L)$ and $%
(-L,-L_{1})$ respectively. Outside the interval $(-L_{1},+L_{1})$ %
\ the ''supercurrent'' $J_{S}$  is zero (this follows directly from 
the expressions for $J_{S}$ and for $\widehat{F}^{R}$ ). As 
there is no current between the N reservoirs, the integrals over energy $%
\epsilon $ from $J_{2+,2-}$ should be equal to zero. In the 
absence of the temperature gradient we obtain from Eq.(5) $f_{-}=0$
and $J_{S}f_{eq}=J_{1eq}=-r(g_{z-}+g_{z+})f_{eq}.$  According Eq.(1) 
the constants $J_{1}$ and $J_{2+,2-}$ are related to each 
other ( Kirchoff's law) 
 
\begin{equation} 
{\Large J}_{2+,2-}-\delta J_{1}=\pm r[g_{1+}f_{-}-g_{z+}\delta f_{+}]_{(\pm 
L_{1})} 
\end{equation} 
 
 where $\delta J_{1}=J_{1}-J_{S}$ and $\delta 
f_{+}=f_{+}-f_{eq}. $ Integrating Eq.(5) outside the interval $%
(-L_{1},+L_{1})$ and taking into account the boundary condition for $%
f_{-}(x)$ , we obtain $J_{2+,2-}=\pm \lbrack f_{-}(\pm L)-f_{-}(\pm 
L_{1})]/L_{2}$. Using this expression and Eq.(6), one can show that \ 
in the main approximation in $r$ the function $f_{-}(x)$ is 
almost constant along the N wire and equal to $f_{-}(x)\approx 
f_{-}(0)\approx f_{-}(\pm L)=eV_{T}\beta \cosh ^{-2}(\beta _{o}\epsilon ).$%
 Integrating Eq.(6) over energies we readily find $eV_{T}$ 
 
\begin{equation} 
eV_{T}=\frac{L_{1}}{L}\frac{\delta T}{T_{o}}\int_{0}^{\infty }d\epsilon  
\text{ }g_{z+}\epsilon \cosh ^{-2}(\beta _{o}\epsilon )/\int_{0}^{\infty 
}d\epsilon \text{ }g_{1+}\cosh ^{-2}(\beta _{o}\epsilon ) 
\end{equation} 
 
The potential $V_{T}$ equals approximately the voltage difference between 
the N reservoirs and superconducting loop. It is worth noting that $V_{T},$ 
determined by Eq.(7), does not depend on the small parameter $r$ \ because 
both functions \ $g_{z+}$\ and $g_{1+}$\ are proportinal to $r$ 
(however one should have in mind that according to the condition (2) this 
parameter must not be too small). The integrand in Eq.(7) can be calculated 
if \ the function $\widehat{F}^{R}$ is known from an approximate or 
numerical solution of the Usadel equation. In the limit considered, of small  
$r$, the retarded (advanced) Green's functions are readily found from the 
linearized Usadel equation. In this case we find 
 
\begin{equation} 
g_{z+}=r%
\mathop{\rm Re}%
(F_{S}^{R}\frac{\sinh ^{2}\theta _{2}}{\theta \sinh (2\theta )})\sin \varphi 
;\text{ \ }g_{1+}=r%
\mathop{\rm Im}%
\{F_{S}^{R}[\sinh (\theta _{2}+2\theta _{1})+\cos \varphi \sinh \theta _{2}]%
\frac{\sinh \theta _{2}}{\theta \sinh (2\theta )}\} 
\end{equation} 
where $\theta =\theta _{1}+i\theta _{2},\theta _{1(2)}=k_{\epsilon 
}L_{1(2)},k_{\epsilon }=\sqrt{-2i\epsilon /\epsilon _{L}}$,$L_{2}=L-L_{1}$ ,$%
F_{S}^{R}=\Delta /\sqrt{(\epsilon +i\Gamma )^{2}-\Delta ^{2}}$ is the 
retarded Green's function in the superconductor. One can see that the ratio 
in Eq.(7) indeed does not depend on $r.$ Numerical analysis of the Usadel 
equation shows that at a characteristic energy $\epsilon \cong \epsilon _{L}$ 
the difference between the linearized and exact numerical solutions of the 
Usadel equation is less than 10\% even for $r\approx 1.$ 
 
It follows from Eqs.(7-8) that the voltage $V_{T}$ caused by the temperature 
gradient is zero when the phase difference between the superconductors is 
zero and oscillates with increasing $\varphi $. One can easily estimates the 
order of magnitude of $V_{T}$. We find 
 
\begin{equation} 
eV_{T}=\delta T(L_{1}/L)\sin \varphi \left\{  
\begin{array}{c} 
(T/\epsilon _{L}){\it C}_{1}(\varphi ),T<<\epsilon _{L} \\  
(\epsilon _{L}/T)^{3/2}{\it C}_{2}(\varphi ),T>>\epsilon _{L} 
\end{array} 
\right. 
\end{equation} 
 
Here ${\it C}_{1,2}(\varphi )$ are periodic functions of the phase 
difference $\varphi $ of order 1 and are not zero when $\varphi =0.$ 
 
If we define the thermoemf $V_{T}$ as the voltage between the N reservoirs 
and the superconducting circuit, we can state that this thermoemf is much 
larger than the thermoemf between the N reservoirs in the absence of 
superconductors because $V_{T}$ does not contain the small parameter $(T$/$%
\epsilon _{F})$ as is the case in a normal system (see for example Ref.\cite 
{Abr}). In addition, $V_{T}$ oscillates with an applied magnetic field $H$ ($%
\varphi $ is proportional to $H$) allowing one to detect small temperature 
gradients. We stress once again that the thermoemf analysed in this 
work arises not between the normal reservoirs, as takes place in case of the 
ordinary thermoelectric effect,  but between the superconducting and 
normal circuits. In Fig.2 we plot the dependence $V_{N}$ on $\varphi $ for 
various $\beta ^{-1}=(2T)$. In Fig.3 the temperature dependence $V_{T}$ at $%
\varphi =\pi /2$ is presented. We see that this dependence is non-monotonic 
with a maximum at $T\approx \epsilon _{L}$ (reentrant behaviour). We note 
that the influence of the proximity effect on the ordinary thermoelectric 
effect was studied theoretically in \cite{Lam}. We ignore this effect 
regarding the thermoelectric current $\alpha \nabla T$ as negligable, 
compared to the effect under consideration. In Ref. \cite{Chandr} the 
thermoelectric voltage was measured in complicated S/N structures which 
differ from the simple structure considered by us. The reentrant behaviour 
of $V_{T}$\ and its oscillations on $\varphi $\ were observed in this work. 
It is possible that the observed effects are related to those considered in 
this paper. It is worthwhile noting that the influence of the ordinary 
thermoelectric current $\alpha \nabla T$  on the Josephson effect was 
studied long ago \cite{Galp}. 
 
The physical explanation of the effect is the following. The temperature 
gradient creates a deviation of the distrubtion function $\delta 
f_{+}=-(\delta n+\delta p)$ from the equilibrium form. On the other hand the 
superconductors do not affect this function because complete Andreev 
reflection conserves the total number of excess electrons and holes. The 
function $\delta f_{+}$\ changes the condensate current flowing across the 
S/N interface. \ If $\delta f_{+}$\ is created by an electrical current 
flowing between the N reserviors, it has the same sign at \ $x=\pm L$\ and 
leads to a change in the Josephson current. In the case considered here the 
function \ $\delta f_{+}$\ has different signs at these points and leads to 
a variation of the condesate current of the same sign at different S/N 
interfaces. Therefore, the potential $V_{T}$ arises in the N wire producing 
a subgap current $rg_{1+}f_{-}(\pm L_{1})$ which compensates the current $%
\delta J_{S}$. 
 
In summary, we have calculated the voltage $V_{T}$ between the 
superconducting and normal curcuits in a S/N mesoscopic structure where the 
normal reservoirs are maintained at different temperatures. This voltage 
arises due to a branch imbalance in the N film and oscillates with varying 
phase difference. Its magnitude does not contain the small parameter $%
(T/\epsilon _{F})$ which is present in normal systems and is of the order $%
\delta T(L_{1}/L)/e$. 
 
We are grateful to the Royal Society and to the EPSRC for their financial 
support. We are grateful to Yu.Gal'perin for his useful comments and 
to V.V.Pavlovskii for his assistance.

\begin{figure}
\vspace{0.3cm}
\hspace{3.2cm}
\epsfxsize=6cm
\epsfysize=5cm
\epsffile{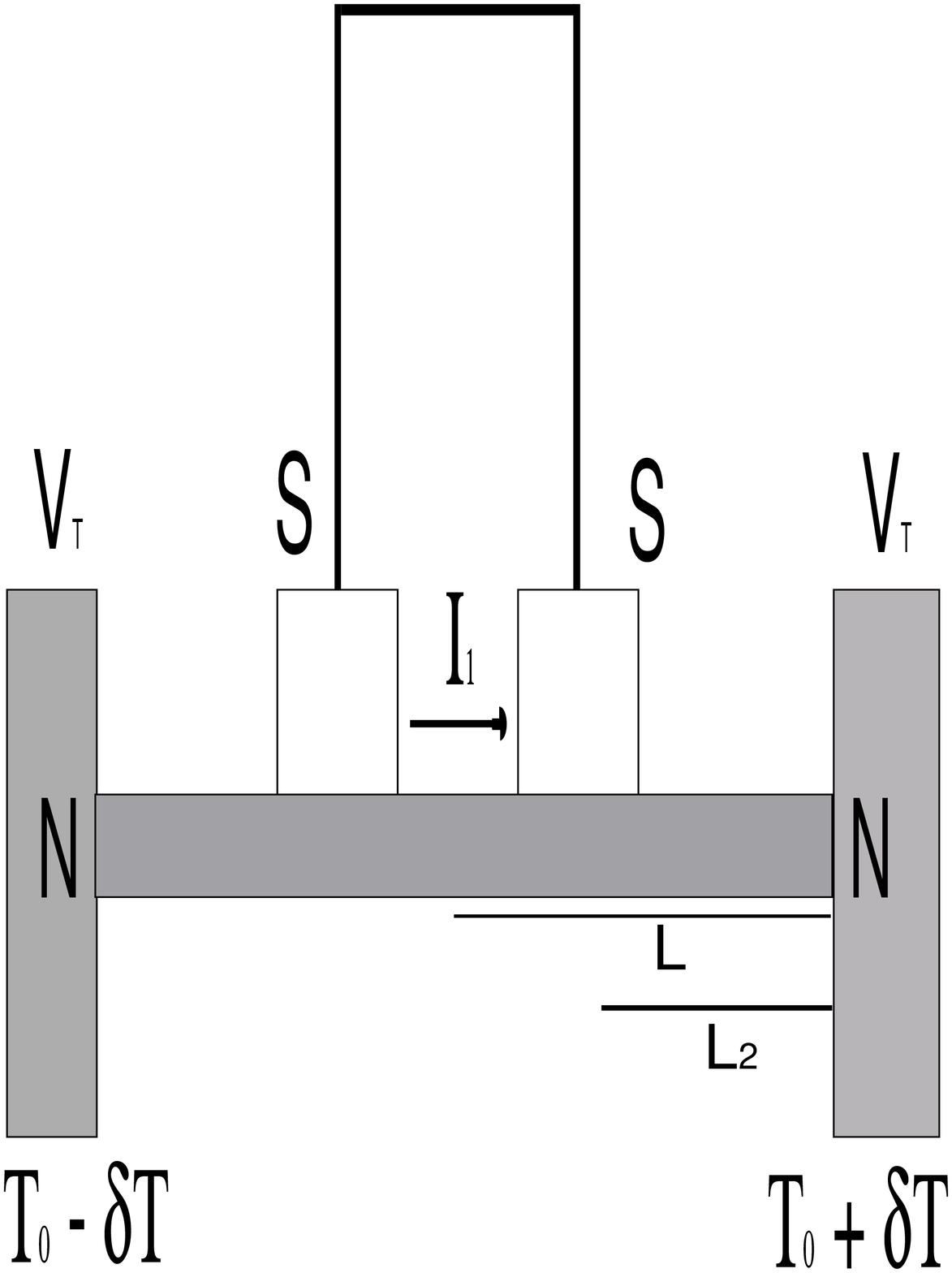}
\vspace{0.3cm}
\refstepcounter{figure}
\label{fig1}
\end{figure}
{\small \setlength{\baselineskip}{10pt} FIG.\ \ref{fig1} Schematic view of 
the 4-terminal S/N/S structure under consideration. The electric potential 
of the superconductors is zero. The N reservoirs are disconnected from the 
external circuit. } 
 
\begin{figure}
\vspace{0.3cm}
\hspace{3.2cm}
\epsfxsize=6cm
\epsfysize=5cm
\epsffile{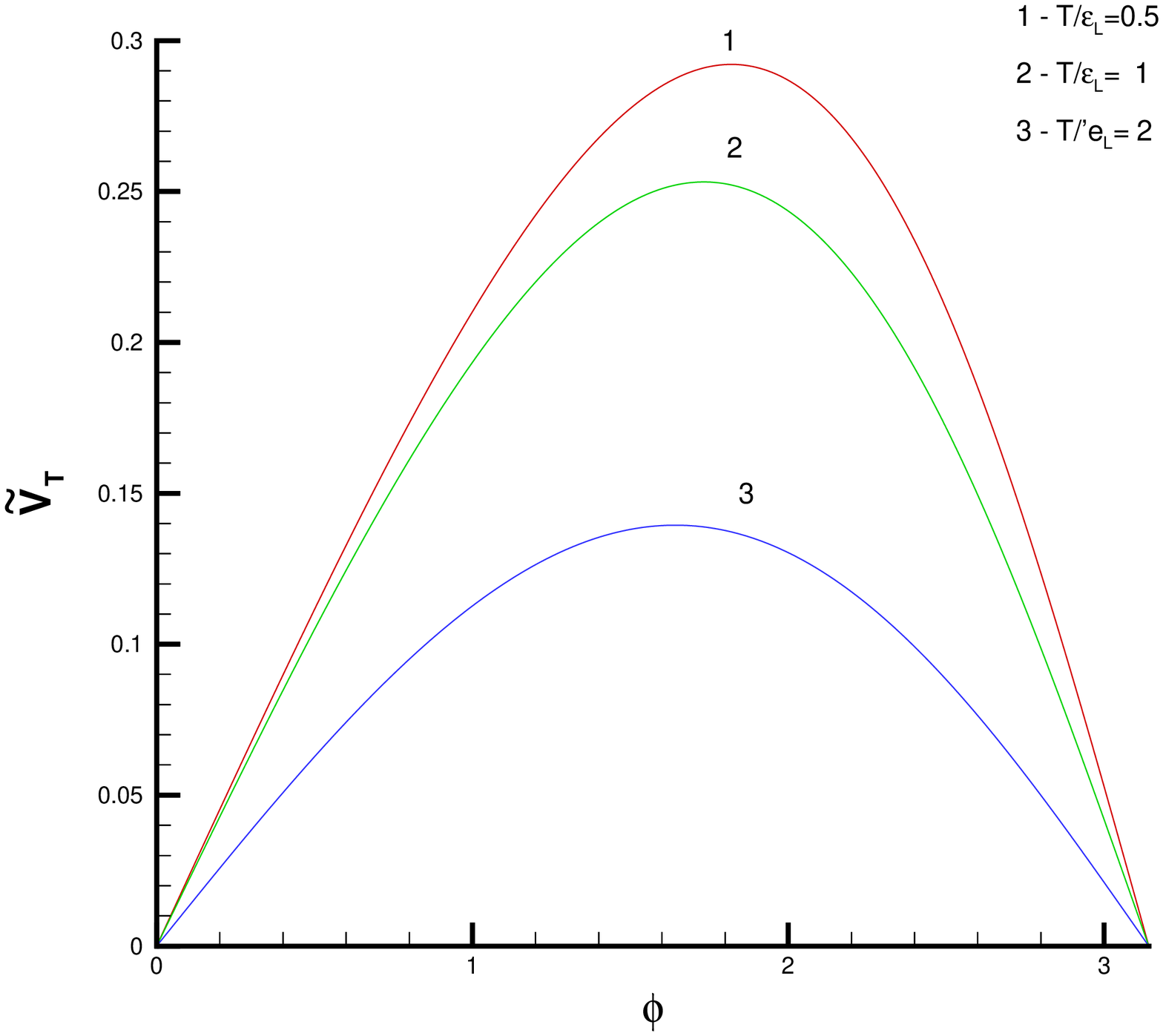}
\vspace{0.3cm}
\refstepcounter{figure}
\label{fig4}
\end{figure}
{\small \setlength{\baselineskip}{10pt} FIG.\ \ref{fig4} The dependence of 
the normalised thermoeletrical voltage $\widetilde{V_{T}}=eV_{T}/(\delta 
TL_{1}/L)$ on the phase difference $\varphi $ for various $\beta =T/\epsilon 
_{L}$  (the parameters are $\Delta /\epsilon _{L}=10,L_{1}/L=0.5,r=0.3$). } 
 
\begin{figure}
\vspace{0.3cm}
\hspace{3.2cm}
\epsfxsize=6cm
\epsfysize=5cm
\epsffile{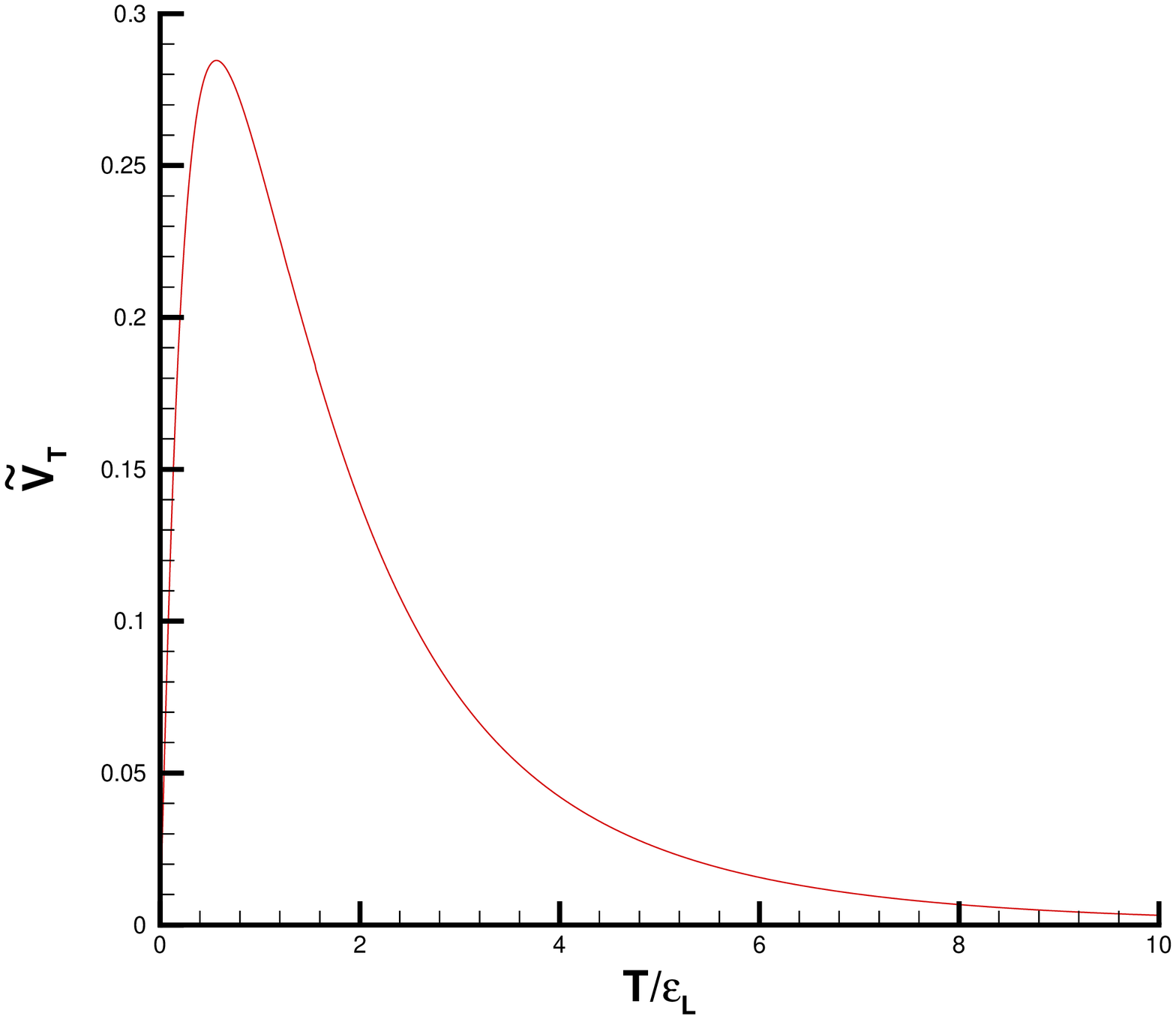}
\vspace{0.3cm}
\refstepcounter{figure}
\label{fig5}
\end{figure}
{\small \setlength{\baselineskip}{10pt} FIG.\ \ref{fig5} The temperature 
dependence of the normalised voltage $\widetilde{V_{T}}$  at $\varphi =\pi /2 
$, for the same parameters as in Fig. 2.} 
 
\end{document}